\documentclass[aps,prl,nofootinbib,
superscriptaddress,twocolumn]{revtex4-2}

\usepackage{ulem} 
\usepackage{amsmath}
\usepackage{amssymb}
\usepackage{bm}
\usepackage{graphicx}
\usepackage{xcolor}
\usepackage{hyperref}

\newcommand{\be}{\begin{equation}}
\newcommand{\ee}{\end{equation}}
\newcommand{\ba}{\begin{eqnarray}}
\newcommand{\ea}{\end{eqnarray}}

\newcommand{\der}[2]{\frac{\partial #1}{\partial #2}}

\def\eps{\varepsilon}
\def\fmc{\rm ~fm^{-3}} 
\def\fm{\rm ~fm}
\def\fm2{\rm ~fm^2}
\def\MeV{\rm ~MeV}    
\def\mL{{\cal L}}
\def\Cs{C_\sigma^2}
\def\Co{C_\omega^2}
\def\Cd{C_\delta^2}
\def\Cr{C_\rho^2}
\def\Lsd{\Lambda_{\sigma\delta}}
\def\Lod{\Lambda_{\omega\delta}}

\begin{document}

\title{Relativistic mean-field model for the ultra-compact low-mass neutron star HESS J1731-347}

\author{Sebastian Kubis}
\email{skubis@pk.edu.pl}
\author{W\l{}odzimierz~W\'ojcik}
\affiliation{Department of Physics, Cracow University of Technology,  Podchor\c{a}\.zych 1, 
30-084 Krak\'ow, Poland}
\author{David Alvarez Castillo}
\affiliation{Henryk Niewodnicza\'nski Institute of Nuclear Physics, Polish Academy of Sciences, Radzikowskiego 152, 31-342 Kraków, Poland}
\author{Noemi Zabari}
\affiliation{Astrotectonics, AstroTectonic Ltd., Słowackiego 24, 35-069 Rzesz{\'o}w, Poland}

\begin{abstract}
The recent observation of the object HESS J1731-347 suggests the existence of a very light and very compact neutron star, which is 
a challenge for commonly used equations of state for dense matter. In this work we present a relativistic mean-field 
model enriched with isovector and isoscalar meson crossing terms. These interactions dominate 
the behavior of the symmetry energy and account for the small radius.  The proposed model fulfills the recent constraints concerning the symmetry energy slope and the state-of-the-art compact star constraints derived from the NICER measurements of the pulsars PSR J0030+0451 and PSR J0740+6620, as well as from the GW170817 event and its associated electromagnetic counterparts AT2017gfo/GRB170817A.
\end{abstract}


\maketitle

\section{Introduction\label{sec:intro}}

The recent observation of a neutron star labelled as HESS J1731-347 \cite{Doroshenko:2022} that has been identified as a very compact and light object challenges most of the state-of-the-art theoretical descriptions. To fulfil constraints from other observations, namely gravitational waves from GW170807 \cite{TheLIGOScientific:2017qsa} and radio/X-ray detection of pulsars among others \cite{Miller:2019cac,Miller:2021qha}, realistic equations of state (EoSs) for compact stars tend to be stiff at higher densities (predicting more massive objects),
whereas at lower densities they should be relatively soft. The characteristic feature of such EoSs is that the mass--radius relation 
has a Z-like shape, with low-mass stars more compact than their high-mass counterparts.
In this work we present a nuclear model  capable of fulfilling all the recent observational constraints, including the inferred mass and radius values of  HESS J1731-347.

HESS J1731-347 has been determined to be a very low-mass, small, compact star, associated with a supernova remnant. This is not the first time an object with such characteristics has been reported. However, in most other cases, corrections due to observations of radiating hot spots instead of the entire surface of the star had to be applied, resulting in larger stellar objects. One of the most important properties of compact stars featuring such surface spots is the signal pulsation. It is indeed the detections of  X-ray  pulsations produced by radiating spots that have allowed the NICER detector to derive estimates of the masses and radii of PSR J0030+0451 and PSR J0740+6620.
The case of HESS J1731-347 has been reported as an exceptional central compact object (CCO), which is isolated, radio-quiet, and non-accreting but thermally radiating. 
No pulsations were detected in its thermal emission, and its atmospheric composition has been determined to be mostly carbon. Moreover, the distance to this source, a very important quantity for electromagnetic flux determination, has been robustly derived from Gaia observations \cite{Bailer-Jones:2021anj}.

The very small stellar radius for light neutron stars considerably reduces the available parameter space for nuclear EoSs  and highlights the potential of the HESS J1731-347 observation. The nuclear symmetry energy, whose stiffness is usually characterized by its slope-related parameter $L$, has a considerable impact on the compact star radius. The general trend is that larger $L$ values result in larger stellar radii. Here we show that a small $L$ around 40 MeV, being in agreement with the results of nuclear experiments,  ensures a compact size for HESS J1731-347.

In the next  section, we introduce the pure hadronic relativistic mean-field (RMF) model with scalar--scalar and scalar--vector meson interactions and provide details of its properties. In the Results section,  the important features of compact stars
are discussed and compared to the recent observational data.

\section{The model\label{sec:model}} 
In the RMF model, the interactions of nucleons are mediated by four types of mesons: $\sigma$, $\omega$, $\rho$, and $\delta$.
The proposed Lagrangian $\mL$ includes $\mL_\text{kin}$, the standard kinetic part for mesons and nucleons; $\mL_{N\phi}$, Yukawa-type couplings of nucleons to mesons; $U(\sigma)$, the
self-interaction term for $\sigma$; and $\mL_\text{cross}$, meson--meson interactions (crossing terms) between $\delta$ and $\sigma$ and between $\delta$ and $\omega$
(further details are presented in \cite{Zabari:2019ukk}):
\begin{equation}
\mL =  \mL_\text{kin} + \mL_{N\phi} - U(\sigma) + \mL_\text{cross} ~,~~ \phi=\sigma,\omega,\rho,\delta .
\label{Ltot}
\end{equation}
The most  crucial ingredient is the crossing term
\begin{equation}
\mL_\text{cross} =
 \frac{1}{2} g_{\sigma\delta}\, \sigma^2\vec{\delta}^2 +  \frac{1}{2}
g_{\omega\delta}\,\omega_\mu\omega^\mu \vec{\delta}^2 ,
\end{equation}
which represents quadratic coupling of two scalar mesons, $\sigma$ and $\delta$, and vector $\omega$ mesons to scalar $\delta$ mesons.
In this paper, we extend the model presented in \cite{Zabari:2019ukk} to $\omega$--$\delta$ coupling.
It is interesting that the model does not include other meson--meson interaction terms, except the very common self-interaction term 
for the $\sigma$ meson, but leads to results in complete agreement with the present set of relevant astrophysical observations.
For future reference, we name the model Cracow Crossing Terms (CCT).
The total Lagrangian $\mL$ leads to the following equations of motion:
\begin{align}
  m_\sigma^2 \bar{\sigma} & =g_\sigma\left(n_p^s+n_n^s\right)-U'(\bar{\sigma}) -  g_{\sigma\delta}
\bar{\sigma}(\bar{\delta}^{(3)})^2 \label{eom1},\\
m_\omega^2{\bar{\omega}}_0 & = g_\omega n - g_{\omega\delta} \bar{\omega}_0 (\bar{\delta}^{(3)})^2  \label{eom2}, \\
m_\rho^2 {\bar{\rho}}_{0}^{(3)} & =\frac{1}{2}g_\rho\left(2x-1\right)n  \label{eom3} ,\\
m_\delta^2 {\bar{\delta}}^{(3)} & = g_\delta\left(n_p^s-n_n^s\right) + g_{\sigma\delta}\bar{\sigma}^2\bar{\delta}^{(3)} +
g_{\omega\delta} \bar{\omega}_0^2 \bar{\delta}^{(3)} \label{eom4},
\end{align}
where the scalar densities are $n_i^s = \frac{2}{(2 \pi)^3}  \int_{0}^{k_{F,i}}  \frac{m_i}{\sqrt{k^2 + m_i^2}} \, d^3 k $.
The mean field of the $\delta$ meson introduces nucleon effective mass splitting \cite{Kubis:1997ew}:
\begin{align}
m_p = m-g_\sigma\bar{\sigma}-g_\delta{\bar{\delta}}^{(3)} ,\label{mp} \\
m_n = m-g_\sigma\bar{\sigma}+g_\delta{\bar{\delta}}^{(3)} . \label{mn}
\end{align}
The effective masses can be  used to replace the meson mean field of nucleons.
Then, after including Eqs.~(\ref{eom1})--(\ref{mn}), the energy density for nucleonic matter
may be expressed in terms of only the densities and the effective nucleon masses:
\begin{align}
  \eps_\text{nuc} = &  \sum_{i=p,n}\frac{1}{4}(3 E_{F,i} n_i + m_i n_i^s) +
   \frac{1}{2 \Cs}(m-\bar{m})^2 + \nonumber \\
 & \frac{\Co}{2} \frac{n^2}{1 + \Co \Lod (\Delta m/2)^2} +
  \frac{\Cr}{8} (2x-1)^2 n^2 +\nonumber \\
 &  \frac{\Delta m^2}{8 \Cd} +  \frac{1}{8}\Lsd (m-\bar{m})^2\Delta m^2 + U(m-\bar{m}),
 \label{enuc}
\end{align}
where the first term in Eq.~(\ref{enuc}) represents the energy of the nucleonic Fermi sea,
i.e., the integrals $\frac{2}{(2 \pi)^3}  \int_{0}^{k_{F,i}}  {\sqrt{k^2 + m_i^2}} \, d^3 k $.
The expression for the energy density shows that it is convenient to replace the coupling appearing in the Lagrangian by  the following  parameters: 
\begin{align}
 & C^2_\sigma=\frac{g_{\sigma}^2}{m_{\sigma}^2},\ C^2_\omega=\frac{g_{\omega}^2}{m_{\omega}^2},\ C^2_\rho=\frac{g_{\rho}^2}{m_{\rho}^2},\ C^2_\delta=\frac{g_{\delta}^2}{m_{\delta}^2} \nonumber ,\\
 & {\rm and} ~~ \Lsd = \frac{g_{\sigma\delta}}{g_\sigma^2 g_\delta^2} ,~~
  \Lod = \frac{g_{\omega\delta}}{g_\omega^2 g_\delta^2} .
   \label{enuc} 
\end{align}
Together with the two constants $b$  and $c$ appearing in the $\sigma$ self-interaction  potential, $U\left(\sigma\right)=\frac{1}{3}b
 m {(g_\sigma\sigma)}^3+\frac{1}{4}c {(g_\sigma\sigma)}^4$, our model possesses eight free parameters: four in the isoscalar sector, $\Cs, \Co, b, c$, and four in the isovector sector,
$\Cr, \Cd, \Lsd, \Lod$. These constants cannot be uniquely fitted to the saturation point properties and hence we have some 
freedom to control the properties of the EoS\@.  One free parameter  from the isoscalar sector, $\Cs$, is used to control the stiffness of the EoS and a second from the isovector sector, $\Cd$, controls the symmetry energy behavior.

The isoscalar constants are fitted to the three features: 
the minimum energy ($\der{\varepsilon/n}{n} =0$) of symmetric matter at $n_0 =0.16 \fmc$ and its compressibility $K_0 = 230 \MeV$.
The fourth one, $\Cs$, being a free parameter, takes a value between 12 and 14 $\fm2$ and ensures that the stiffness of
the EoS is sufficient to produce a stellar mass in agreement with observational limits. 
According to recent observations, the maximum neutron star mass must be above $2.1 M_\odot$, and the adopted values of $\Cs$
allow for such stellar masses.
\begin{table}[t!]
\def\szer{1.2cm}
\caption{Coupling constants for nine CCT models numbered by the values of  $\Cs$ and  $L$.}
\begin{tabular}{|c|ccc|c|cc|}
\hline
$\Cs$ &  $\Co$ & $b$ & $c$ & $L$ & $\Cr$ & $\Cd$ \\ 
 $ \fm2 $& $\fm2$ & - & - & MeV & $\fm2$ & $\fm2$ \\ \hline
     &   &  &                          & 40 & 15.2938 ~ &  2.63799 ~ \\
 12  & 6.9769 ~  & 0.004733 & ~$-$0.0052878& 60 & 13.1107 & 2.24899 \\
     &  &  &                           & 80& 9.93091 & 1.61518 \\ \hline
     &   &  &                          & 40 & 13.9852 & 2.33674 \\
 13  & 7.9531 ~  & 0.003695 & ~$-$0.0045224& 60 & 12.1548 & 2.02896 \\
     &   &  &                          & 80 & 9.60636 & 1.54906 \\ \hline
     &  &  &                           & 40 & 12.9888 & 2.11461 \\
 14  & 8.9055 ~  & 0.003005 & ~$-$0.0039370& 60 & 11.4361 & 1.86945 \\
     &  &  &                           & 80 & 9.36054 & 1.50265 \\ \hline
\end{tabular}
\label{tab-const}
\end{table}
At the isovector sector, the four constants have to be determined, but only the two 
saturation point properties  are available: the symmetry energy $E_\text{sym}(n_0)$ and its slope $L(n_0) = 3 n_0 \frac{dE_\text{sym}}{dn}$.

In the work \cite{Zabari:2019ukk}, the $\sigma$--$\delta$ meson crossing term was introduced and it was shown that it leads to the 
desired soft symmetry energy. The effects of this term on the neutron star properties were further analyzed in 
\cite{Kubis:2020ysv}. The model with a $\sigma$--$\delta$ meson crossing term was also extended to the other meson crossing term, as was done in \cite{Miyatsu:2022wuy}.

 However, the effect of $\Lsd$ on the $E_\text{sym}$ softening occurs only in the
 vicinity of the $n_0$ as the  $\sigma$ meson mean field contributes less to the total energy when the density increases.
 At higher densities, the vector meson fields are those that contribute more to the total energy.
 That was our motivation to additionally couple the scalar  meson $\delta$  to the vector meson  $\omega$ and test whether such coupling is in agreement with dense matter properties and astrophysical observations. 

 Then the expression for the  symmetry energy is  
\begin{widetext}
\begin{equation}
  E_\text{sym}(n) = \frac{1}{8}\Cr n + \frac{k_0^2}{3 E_{F,0}} - 
  \Cd\,\frac{m_0^2 n}{2 E_{F,0}^2(1 - 3 \Cd(\frac{n}{E_{F,0}}-\frac{n_s}{m_0}) - \Cd\Lsd (m-m_0)^2 - \Co\Lod n^2 )},
\end{equation}
\end{widetext}
where $E_{F,0}$, $m_0$, and $n^s = n_p^s + n_n^s $ are the Fermi energy, effective nucleon mass, and scalar density in symmetric matter.
By the presence of the two constants $\Lsd$ and $\Lod$ in the denominator, one may reduce the slope of the symmetry energy for a 
wider range of densities.
 Both these  couplings are fixed at $\Lsd = 0.032$ and $\Lod = 0.02$ and  the remaining  isovector  constants $\Cr, \Cd$ 
 are used  to  obtain the required values of $E_\text{sym}(n_0)$ and $L$. 
 
 It is commonly accepted that $E_\text{sym}(n_0) = 30 \rm \MeV$; however, the value of $L$ is still a matter of discussion.
Different nuclear  experiments give values spread across a wide range. Those based on the spectral pion ratio suggested $47 < L < 117 \MeV$
\cite{SpiRIT:2021gtq}.
The most promising experiments based on the measurement of neutron skin, which is correlated with the slope value, give different 
results.
Analysis from PREX-2 ($\rm ^{208}Pb$ nucleus) \cite{PREX:2021umo,Reed:2021nqk} suggested $L=106 \pm 37 \MeV$. 
Including the correlation 
with  parity-violating asymmetry for $\rm ^{208}Pb$ gives $L=54\pm 8\MeV$ \cite{Reinhard:2021utv}. 
Experiments with a lighter nucleus, $\rm ^{48}Ca$, give even smaller values for $L$. The authors of \cite{Reinhard:2022inh}
presented a combined analysis of $\rm ^{208}Pb$ and $\rm ^{48}Ca$ and, in the framework of DFT functionals,
were not able to reconcile this discrepancy. They suggested that $L$ is in the range from 15 to 83 MeV.
In the work \cite{Lynch:2021xkq}, the authors point out that the different results for $L$ come from the fact that
different experiments probe the  symmetry energy at different densities.
Recently, Lattimer \cite{Lattimer:2023rpe} extensively discussed the sources of observed discrepancies in the slope 
measurement  and suggested that the most likely value is between 40 and 50 MeV.
In view of the above difficulties, we suggest adopting values of $L$ from the range from 40 to 80 MeV.

To sum up, we set up a two-parameter family of nuclear models, described by the $\Cs$  coupling and the value of the slope $L$.
The values of all relevant coupling constants are given in Table~\ref{tab-const}.

\section{Results\label{sec:results}}

\begin{figure}[h!]
	\begin{center}
		\includegraphics[width=1.1\columnwidth]{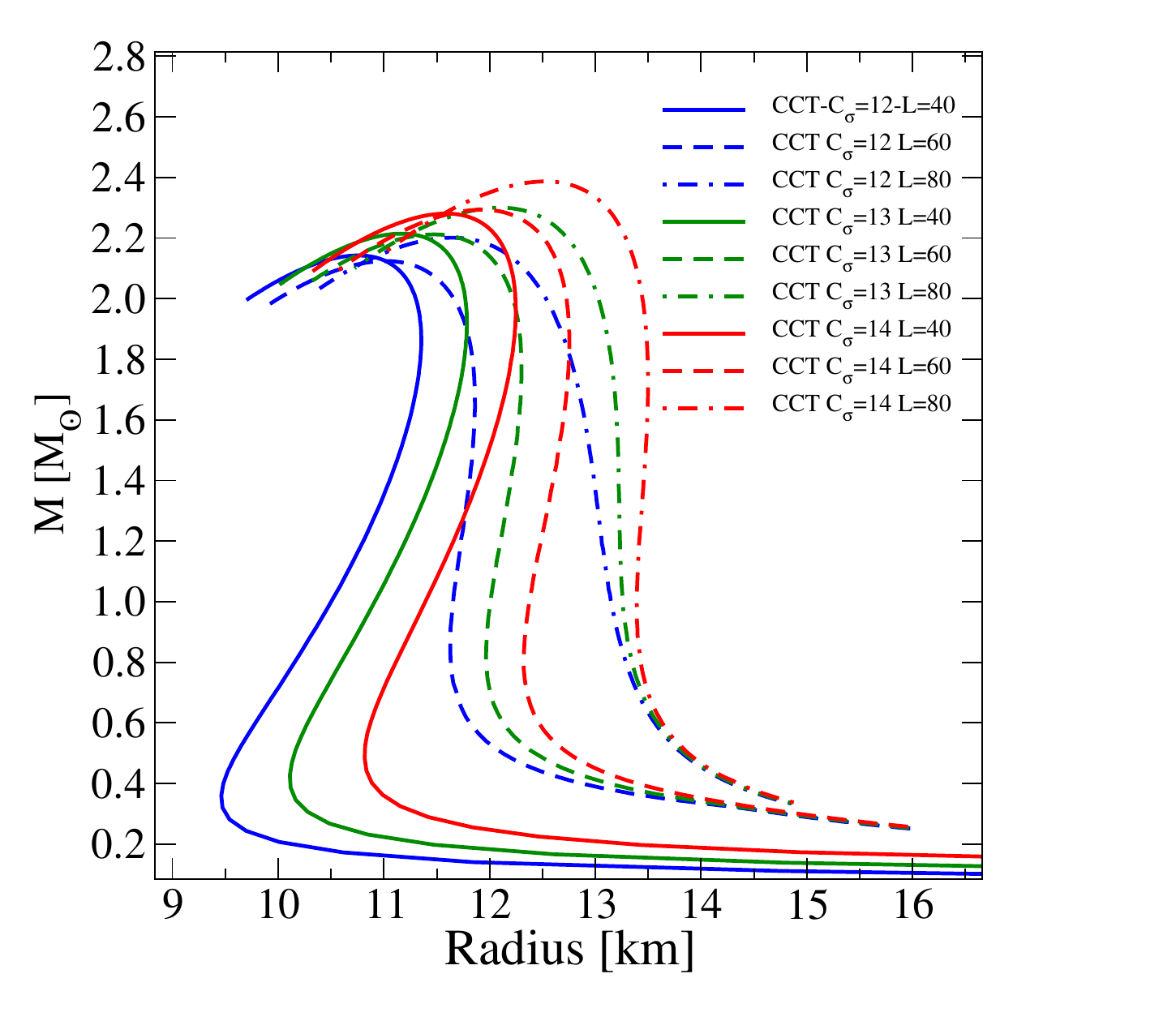}\\
		\includegraphics[width=1.1\columnwidth]{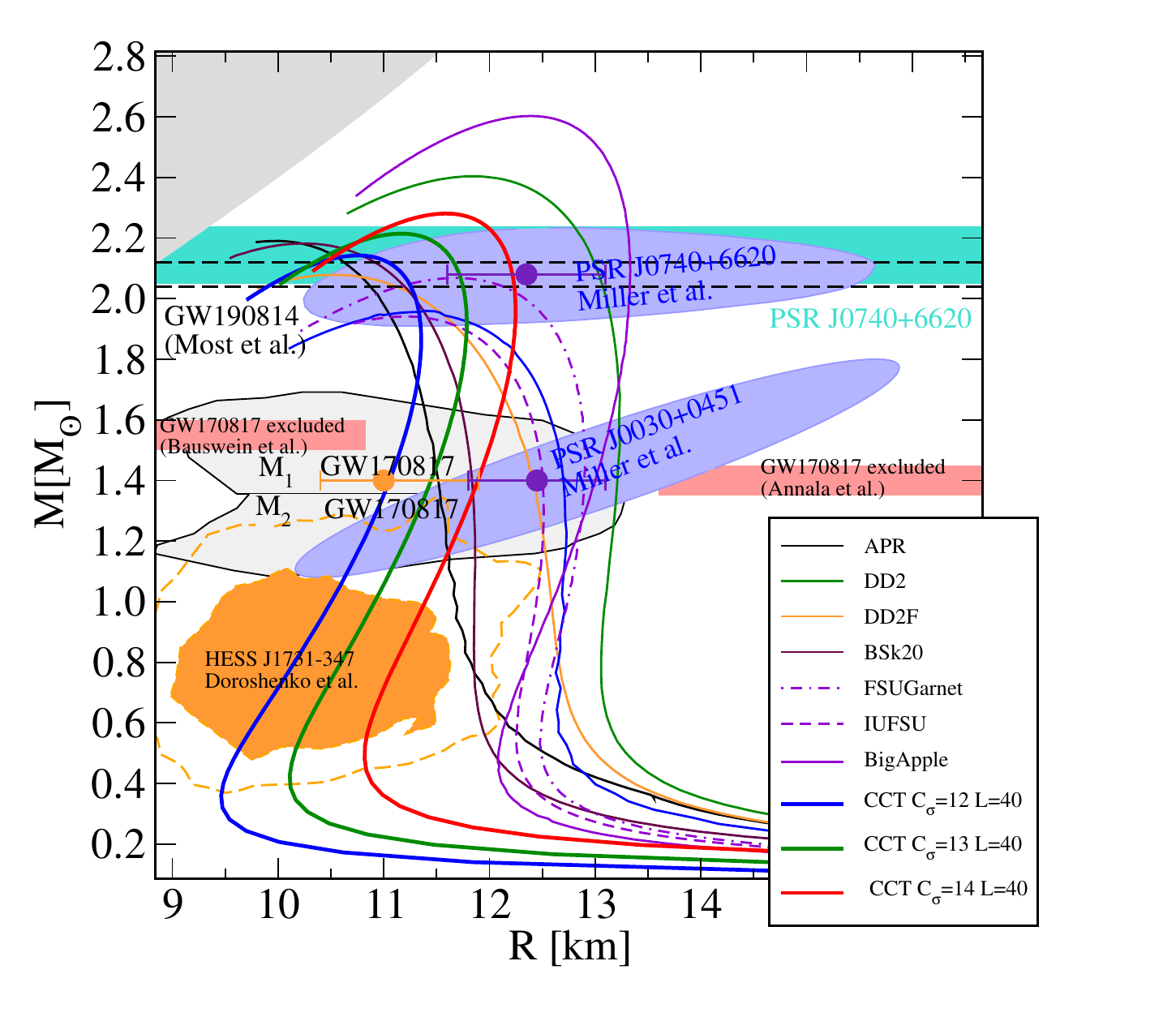}
	\end{center}
	\caption{\label{Fig:MR}
		Mass--radius diagrams of compact stars. The upper panel presents the results for the CCT RMF models introduced in this work, with the  Z-like curves obtained for the lowest values of $L$. The lower panel presents the models with the lowest $L=40 \MeV$ and three $\Cs$ values. The curves from other EoS approaches are included. Observational constraints are shown as colored regions and points with error bars; see the text for details.}
\end{figure}

The introduced RMF model was used to derive the properties of a spherically symmetric configuration of a neutron star. 
The beta equilibrium applied to dense matter with leptons allows finding the equation of state in the core region,
whereas for the crust we applied the Sly4 model \cite{Chabanat:1997un,Douchin:2001sv}.
For the model with the lowest  $L=40  \MeV$, the phase transition occurs and the Maxwell construction was used to
describe the phase coexistence. The mass--radius diagram for compact stars is shown in Fig.~\ref{Fig:MR}, where the upper panel shows the curves for the full parameter space, characterized by the coupling constant $C_{\sigma}$ and the $L$ parameter, which are depicted by various line styles  and colors, respectively.  The lower panel includes
the three chosen curves with $L=40 \MeV$ and different $\Cs$ values,
and also stellar sequences from other EoSs. Additional EoSs in the lower panel include APR~\cite{Akmal:1998cf}, DD2, DD2F~\cite{Typel:2014tqa}, BSK20~\cite{Chamel:2011aa}, FSUGarnet, IUFSU, and BigApple~\cite{Fattoyev:2020cws}. The latter three ones are similar RMF models to the one we have introduced in this work.
In the lower panel, the present compact star measurements and  constraints derived from astrophysical observations are shown. 
Compact star measurements include the observation of the object
labelled as HESS J1731-347 that suggests a very compact object of low mass. The blue elliptical regions correspond to NICER measurements of the objects PSR 7040+6620~\cite{Miller:2021qha} and PSR J0030+0451~\cite{Miller:2019cac}. The violet dots with associated error bars correspond to the derived radius value from an updated analysis of these objects
~\cite{Miller:2021qha}, whereas the orange dot with error bars at 1.4 M$_{\odot}$ is the result of a Bayesian analysis performed while taking into consideration several compact star measurements~\cite{Capano:2019eae}. The green and gray regions correspond to the two estimated masses of the components of the binary system that merged and produced gravitational waves in the event GW170817. The upper dashed lines mark a lower-bound interval for the lower component of the event GW190814 under the assumption that it was a rapidly spinning neutron star~\cite{Most:2020bba}. Red regions correspond to estimated excluded regions from the analysis of the GW170817 event 
\cite{Bauswein:2017vtn,Annala:2017llu}.

\begin{figure}[htpb!]
		\includegraphics[width=1.0\columnwidth]{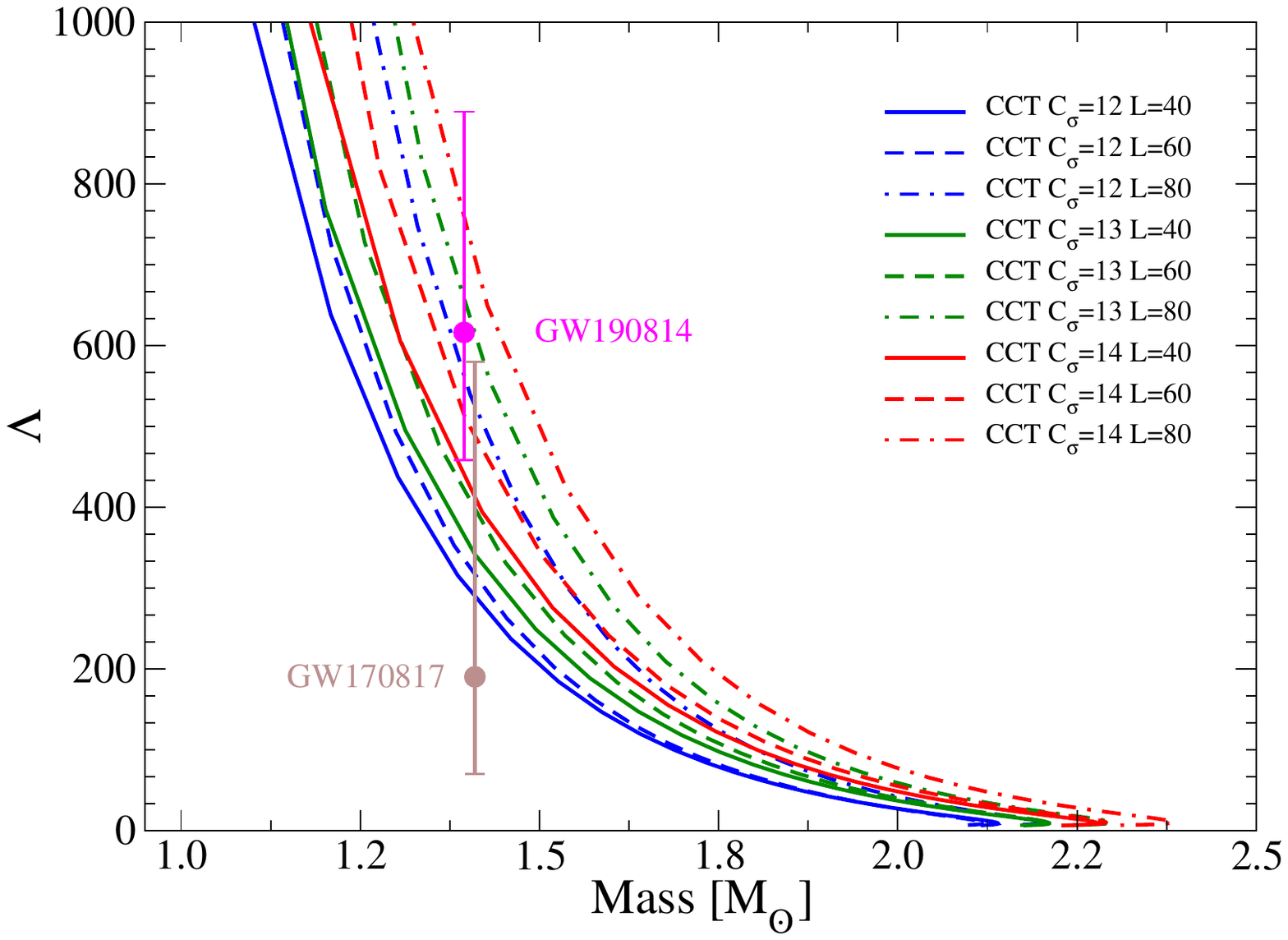}\\
		\caption{\label{Fig:tidal_deformabilitiesML} The tidal deformability as function of neutron star mass. Points with error bars correspond to the star with mass 1.4 $M_\odot$ for the two
merger events.}
\end{figure}

\begin{figure}[b!]
		\includegraphics[width=1.0\columnwidth]{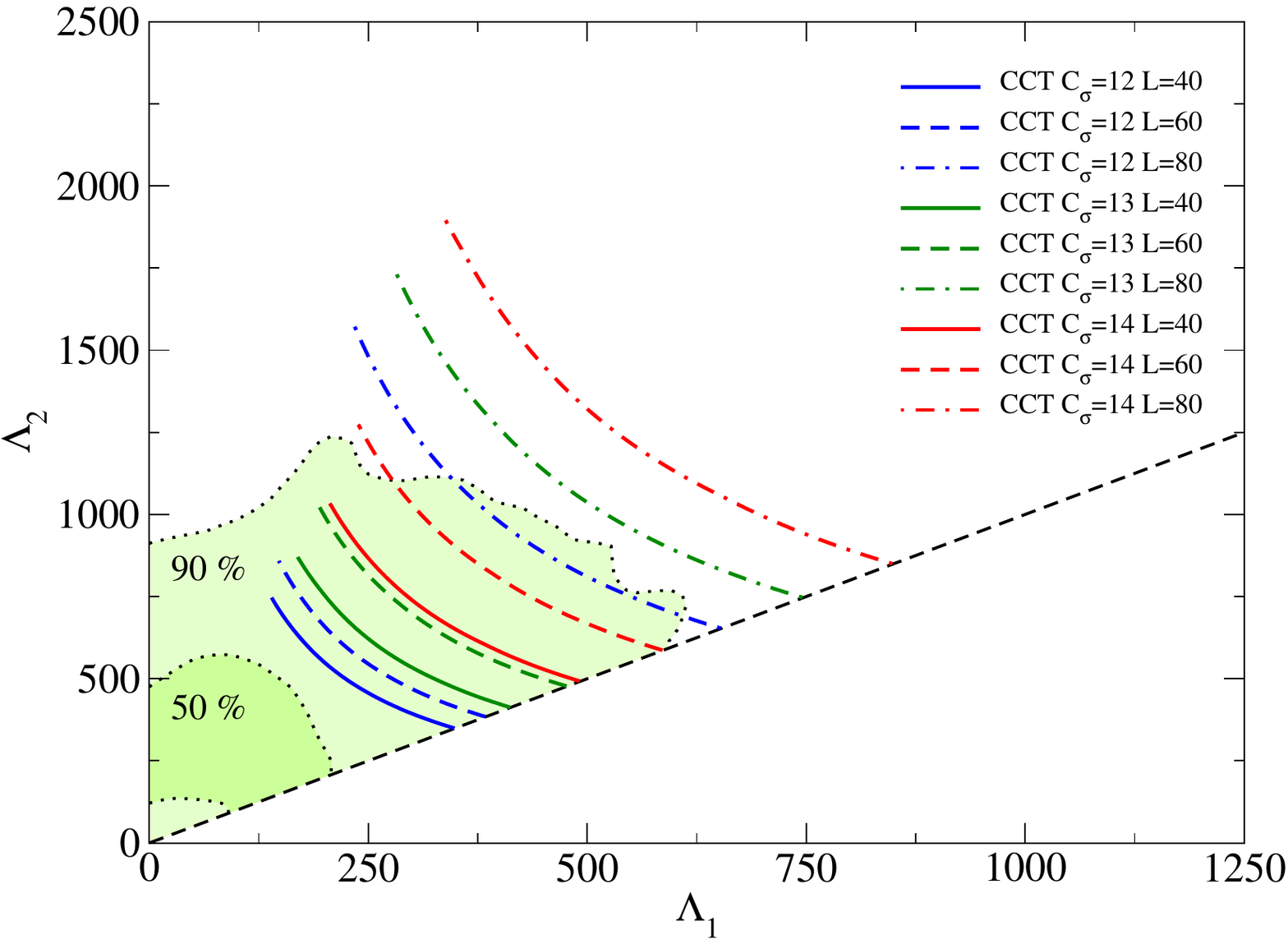}
	\caption{\label{Fig:tidal_deformabilitiesLL}
		Tidal deformabilities for two binary components derived from GW170817. RMF models with lower $L$  better fit to the indicated confidence levels.}
\end{figure}

Figures \ref{Fig:tidal_deformabilitiesML} and \ref{Fig:tidal_deformabilitiesLL} show tidal deformability values $\Lambda$ for compact star sequences. They are derived following the approach introduced in~\cite{Hinderer:2007mb}; see~\cite{Paschalidis:2017qmb} for a discussion on the implications for the compact star EoS\@. Figure~\ref{Fig:tidal_deformabilitiesML} shows the dependence of the deformability $\Lambda$ on the stellar mass $M$, which obeys the relation
$\frac{2}{3}k_{2}\frac{R^{5}}{M{5}}$, where $k_{2}$ is the stellar Love number. The measurement point with associated error bar in the vicinity of the 1.4 M$_{\odot}$ star is the stellar value estimated from the gravitational wave signal of the neutron star merger GW170817, under the assumption that the neutron stars were slowly spinning~\cite{LIGOScientific:2018cki}. A second estimate from an analysis of the GW190814 event that includes the information from GW170817~\cite{LIGOScientific:2020zkf} is also shown. 
Figure~\ref{Fig:tidal_deformabilitiesLL} displays tidal deformability relations for the two stars participating in GW170817. The green regions correspond to the 90\% and 50\% confidence levels derived from the Bayesian analysis performed by the LIGO-Virgo collaboration ~\cite{LIGOScientific:2018cki}. The general tendency is that more compact stars (which have a lower symmetry energy slope $L$) agree with this measurement better than larger ones.

\section{Conclusions and outlook} 
 
The detection of the compact star associated with the  HESS J1731-347 observation has provided a candidate for an ultra-compact object of low mass. This measurement represents both a challenge and a strong constraint on EoSs for compact stars. It has already been shown in the reporting article~\cite{Doroshenko:2022} that a set of EoSs derived from chiral perturbation theory as well as strange star models~\cite{DiClemente:2022wqp} are able to describe such an ultra-compact star while simultaneously fulfilling state-of-the-art compact star constraints from other observations.

Thereportng article ~\cite{Doroshenko:2022} induced hot discussion about dense matter models being able to explain the very compact object.   Recent works  derived from RMF with density-dependent couplings and tensor forces that are barely compatible with  HESS J1731-347~\cite{Huang:2023vhy} The strange star  models proposed in ~\cite{DiClemente:2022wqp,Das:2023qej} and stars with hyperon content \cite{Li:2023vso}are better in describing this measurement.

In contrast, in this work we have introduced a hadronic RMF with crossing terms among isovector and isoscalar mesons that is able to describe HESS J1731-347 while also reproducing laboratory data such as the saturation properties of nuclear matter. We have found that low values of the symmetry energy slope $L$, around
 $40 \MeV$, for each of the chosen values of the coupling $\Cs$ are best able to reproduce the compact object in HESS J1731-347. The $\Cs$ coupling mainly controls the stiffness of the EoS for more massive stars and the highest value $\Cs = 14 \fm2$ seems to be the most preferable  in view of the  lower bound for maximum mass from precise Shapiro delay measurements for PSR J0740+6620.
 
 The higher stiffness of the EoS and the higher $\Cs$ are also supported by the GW events.
As can be seen from the measurement shown as a violet dot with associated error bar for the 1.4~M$_{\odot}$ star in Fig.~\ref{Fig:MR}, the coupling $C_{\sigma}=14$ fm$^{2}$ (red line) is the best value to fulfil this constraint.
Furthermore, from the $\Lambda$ vs. $M$ relation shown in Fig.~\ref{Fig:tidal_deformabilitiesML}, we can see that the same $C_{\sigma}$ value would best fulfil the shared region of $\Lambda$ for events GW170817 and GW190814, provided the lower mass component 
in GW190814 is indeed a compact star. 

Other properties, such as the cooling features (analyzed in \cite{Sagun:2023rzp})  and rotational configurations of compact stars under the CCT RMF approach presented in this work, are left for a follow-up study.

\section{Acknowledgements}
The authors would like to thank Jorge Piekarewicz for sharing with us a set of equations of state. D.E.A.C. also acknowledges support from the NCN OPUS Project No. 2018/29/B/ST2/02576.

\end{document}